\documentclass[aps,prl,twocolumn,superscriptaddress]{revtex4}
\usepackage{natbib}
\usepackage{amsmath}
\usepackage{graphicx}
\usepackage{dcolumn}
\usepackage{tabularx}
\usepackage{keyval}
\usepackage{multirow}

\begin{document}
\title{Existence, character and origin of surface-related bands in the high temperature iron pnictide superconductor BaFe$_{2-x}$Co$_{x}$As$_{2}$.}
\author{Erik van Heumen}\thanks{E.v.H, J.V. and K.K. contributed equally to the results presented here.}
\email{e.vanheumen@uva.nl}
\affiliation{van der Waals - Zeeman institute, University of Amsterdam, 1018 XL Amsterdam, the Netherlands}
\author{Johannes Vuorinen}
\affiliation{Department of Physics, Tampere University of Technology, PO Box 692, FIN-33101 Tampere, Finland}
\author{Klaus Koepernik}
\affiliation{Institute for Theoretical Solid State Physics, IFW Dresden, D-01171 Dresden, Germany}
\author{Freek Massee}
\affiliation{van der Waals - Zeeman institute, University of Amsterdam, 1018 XL Amsterdam, the Netherlands}
\author{Yingkai Huang}
\affiliation{van der Waals - Zeeman institute, University of Amsterdam, 1018 XL Amsterdam, the Netherlands}
\author{Ming Shi}
\affiliation{Paul Scherrer Institut, Swiss Light Source, 5232 Villigen, Switzerland}
\author{Jesse Klei}
\affiliation{van der Waals - Zeeman institute, University of Amsterdam, 1018 XL Amsterdam, the Netherlands}
\author{Jeroen Goedkoop}
\affiliation{van der Waals - Zeeman institute, University of Amsterdam, 1018 XL Amsterdam, the Netherlands}
\author{Matti Lindroos}
\affiliation{Department of Physics, Tampere University of Technology, PO Box 692, FIN-33101 Tampere, Finland}
\author{Jeroen van den Brink}
\affiliation{Institute for Theoretical Solid State Physics, IFW Dresden, D-01171 Dresden, Germany}
\author{Mark S. Golden}
\affiliation{van der Waals - Zeeman institute, University of Amsterdam, 1018 XL Amsterdam, the Netherlands}
\date{\today}

\begin{abstract}
Low energy electron diffraction (LEED) experiments, LEED simulations and finite slab density functional calculations are combined to study the cleavage surface of Co doped BaFe$_{2-x}$Co$_x$As$_2$ (x = 0.1, 0.17). We demonstrate that the energy dependence of the LEED data can only be understood from a terminating 1/2 Ba layer accompanied by distortions of the underlying As-Fe$_2$-As block. As a result, surface related Fe $3d$ states are present in the electronic structure, which we identify in angle resolved photoemission experiments. The close proximity of the surface-related states to the bulk bands inevitably leads to broadening of the ARPES signals, which excludes the use of the BaFe$_{2-x}$Co$_x$As$_2$ system for accurate determination of self-energies using ARPES.
\end{abstract}
\maketitle

The recently discovered iron pnictide superconductors \cite{kamihara-JACS-2008} have quickly turned into one of the most widely studied systems in condensed matter research, using all the modern tools that have been perfected through years of research on another class of high temperature  superconductors: the cuprates. However, two of the most important and widely used tools to determine the electronic structure, scanning tunneling microscopy (STM) and angle resolved photoemission (ARPES) are surface sensitive techniques. It is therefore crucial to assess the impact of the surface on the electronic structure. 

Several different structures containing As-Fe$_2$-As blocks are known: LnOFeAs, MFeAs and MFe$_{2}$As$_{2}$ commonly referred to as the "1111", "111" and "122" families. Subdividing the unit cell into As-Fe$_{2}$-As blocks and spacer layers results in an alternating stack of positively and negatively charged layers. Because of the alternating charges of these layers, the presence of a surface can give rise to a diverging electric potential in the solid \cite{tasker-JPSSP-1979,ohtomo-NAT-2004}. Such polar surfaces are energetically unfavorable and usually a structural or electronic reconstruction takes place to prevent their occurrence. LaOFeAs is such an example: cleavage takes place  between the LaO and As-Fe$_2$-As blocks, resulting in either a highly reconstructed La or As termination \cite{eschrig-PRB-2010} which supports surface-related states that have been observed in ARPES \cite{liu-PRB-2010}. The surface of the 111 system, on the other hand, barely influences the electronic structure \cite{lankau-arxiv-2010}.

This leaves the MFe$_{2}$As$_{2}$ pnictide superconductors which are by far the most widely studied using surface sensitive probes. In this case, STM experiments show a diversity of topographies resulting from the cleaving process \cite{hsieh-arxiv-2008,yin-PRL-2009, nascimento-PRL-2009, massee-PRB-2009, zhang-PRB-2010,niestemski-arxiv-2009}, which have been interpreted as either consisting of As or Ba layers. Recent first principle calculations for the parent compounds show a reconstructed, Ba terminated surface to be energetically favorable, leaving open the possibility for a metastable As termination upon low-T cleavage  \cite{gao-PRB-2010}. There is a paucity of well-founded experimental data on the exact surface structure of the superconducting 122 compounds, in particular when cleaved at cryogenic temperatures as is usual for ARPES and STS studies. Consequently, the impact of the real surface structure on the electronic states responsible for superconductivity remains an open and important issue.    

In this Letter we combine IV-LEED experiments and simulations with DFT slab calculations to investigate the cleavage surface of superconducting, Co-doped Ba122. The data show that the half Ba-layer termination has an impact on the underlying As-Fe$_{2}$-As block, leading to significant departures from the bulk structure. This gives rise to surface-related Fe $3d$ bands, which, in turn, can be found in our ARPES data. These surface states impede accurate quantitative analysis of (for example the widths of) ARPES features in the Ba122 systems.

\begin{figure}[t]
\centering
\includegraphics[width=8.6cm]{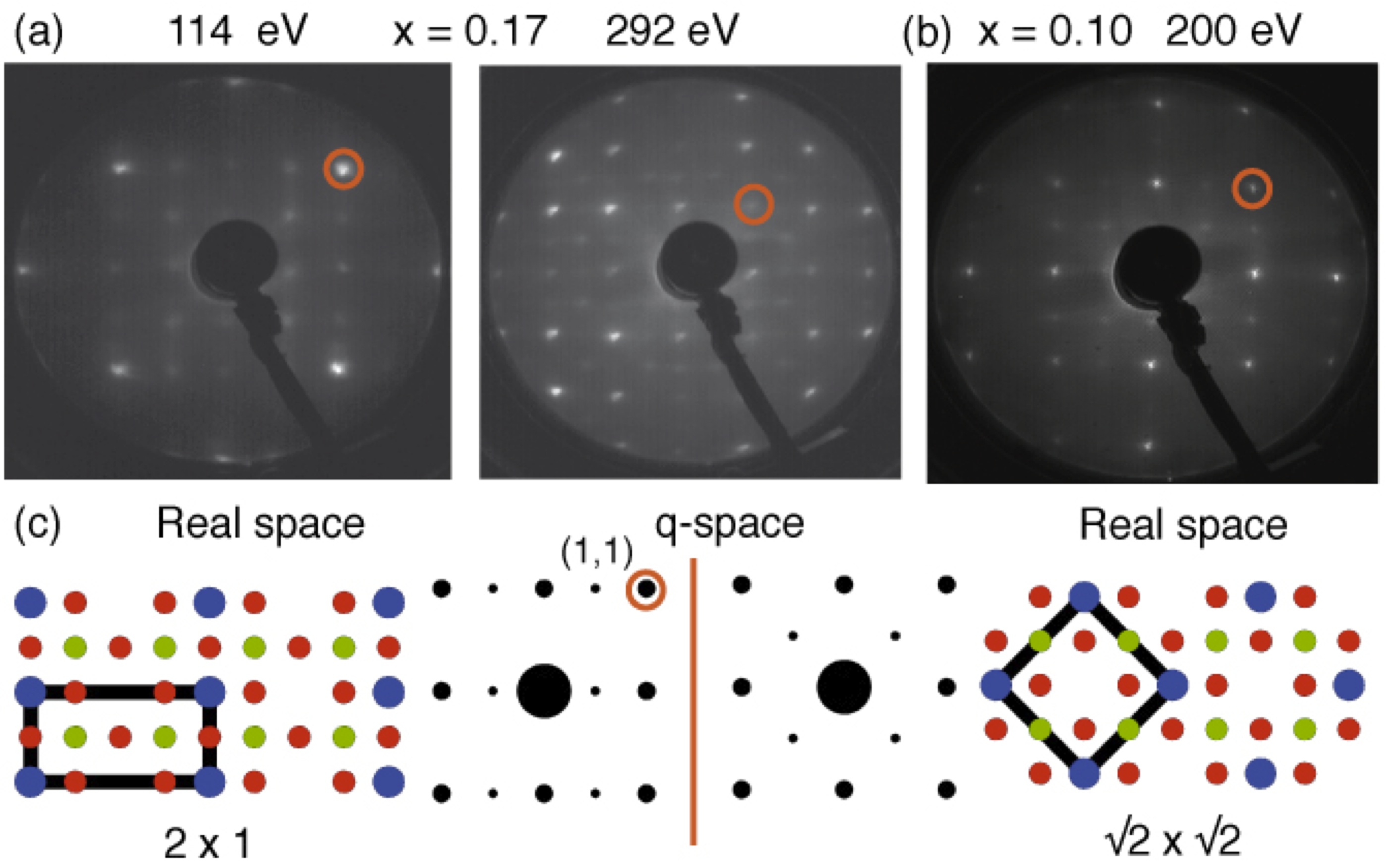}
\caption{(Color online) (a): Typical LEED images for E$_{0}$ = 114 eV and 292 eV for  BaFe$_{2-x}$Co$_x$As$_2$ with x = 0.17. (b) E$_{0}$ = 200 eV for x = 0.1. In each image, representative (1,1) spots are highlighted with red circles. All images taken at 17 K. (c): real space and q-space structures of the $2\times1$ and $\sqrt(2)\times\sqrt(2)$ structures (green: Ba; purple: As; orange: Fe)}
\label{structure}
\end{figure} 
We start with the LEED experiments.  Samples were top-post cleaved  \textit{in-situ} at pressures better than 5$\times10^{-11}$ mbar at temperatures under 20K - following the standard recipes used in ARPES and/or STS studies. For details on samples and experimental procedures, see supplementary materials.        
The measured LEED patterns for x = 0.1 and 0.17 samples [Fig. \ref{structure}(a), \ref{structure}(b)] resemble a $2\times2$ pattern, which in fact arises from a combination of ordered surface structures as indicated in Fig. \ref{structure}(c). In order to circumvent the polar surface (type III in the classification of Ref. \cite{tasker-JPSSP-1979}), half of the Ba layer goes with each half of the cleaved crystal. This now places a terminating 1/2 Ba layer with charge +1 on top of the underlying As-Fe$_2$-As block which has overall charge -2. We obtain a zero net dipole moment if we think of the Ba layer below the uppermost As-Fe$_2$-As block as half belonging to this upper block and half belonging to the next As-Fe$_2$-As block below. Continuing in this fashion represents the only energetically favorable situation of any likelihood \cite{ftnote}. Ordering in the partial Ba termination layer gives rise to the additional, fractional spots in the LEED images as indicated in Fig. \ref{structure}(c). The patterns observed are identified as $2\times1$, $1\times2$ and $\sqrt{2}\times\sqrt{2}$ ordered partial Ba termination layers. The integer $1\times1$ pattern contains higher order reflections from all types of superstructure. The reconstructions are not limited to the tetragonal phase only: the LEED pattern of the underdoped, orthorhombic phase is identical, as shown in Fig. \ref{structure}(b).

\begin{figure}[t]
\includegraphics[width=8.6cm]{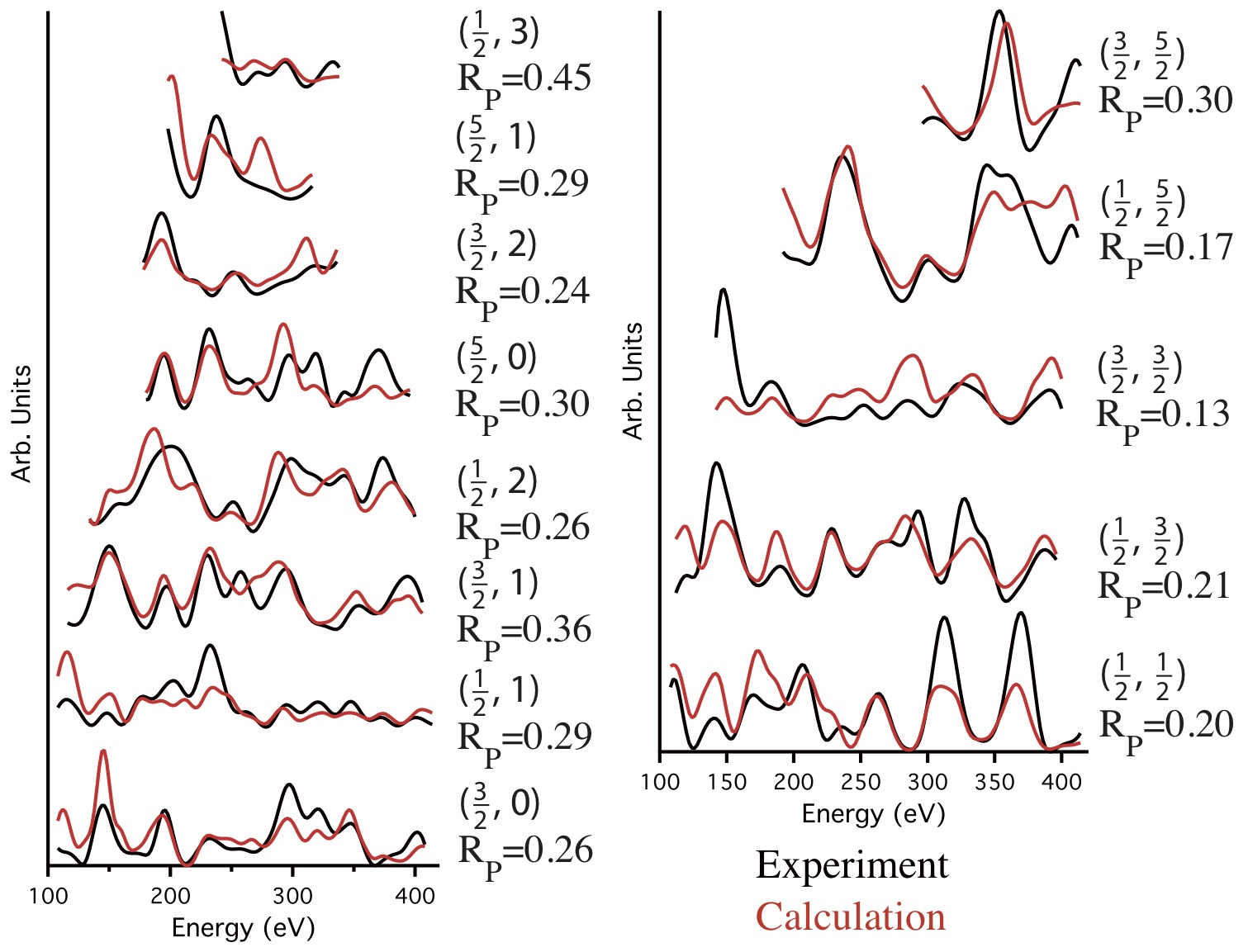}
\caption{(Color online) Half integer IV-curves for both terminations. Left: $2\times1$ termination, right: $\sqrt{2}\times\sqrt{2}$ termination. The Pendry R factor is indicated for each spot.}
\label{IVcurves}
\end{figure}

Even though the arguments presented above on the basis of energetics are already persuasive,we now prove the 1/2 Ba termination using IV-LEED data, in which we solve for the real surface structure of the outermost four atomic layers. Figure \ref{IVcurves} presents the results from an x = 0.17 sample, together with the relevant simulations, carried out using the Barbieri/Van Hove SATLEED package \cite{satleed} (see supplementary material for details). Pendry R-factors \cite{pendry-JPC-1980} are used to measure the level of agreement between experiment and simulation. 
Simulations assuming a bulk crystal terminated by a 1/2 Ba layer with either $\sqrt{2}\times\sqrt{2}$ or $2\times1$ superstructure resulted in overall Pendry R-factors, R$_{P}\approx$ 0.31 and R$_{P}\approx$ 0.46, respectively. As a satisfactory solution should have $R_p$ $<$ 0.3, the ordered superstructures of the half-Ba layer alone are not sufficient. We therefore allowed the atomic positions of the deeper lying layers to relax. Given the measured energy ranges (in total 1208 eV for the $\sqrt{2}\times\sqrt{2}$ and 1782 eV for the $2\times1$ superstructure), optimizing the parameters of the first four atomic layers so as to give reliable estimates of the actual deviations from bulk positions proved to be feasible. We arrive at final Pendry R-factors of R$_{p}$ = 0.19 for the resulting distorted $\sqrt{2}\times\sqrt{2}$ termination and R$_{p}$ = 0.29 for the distorted $2\times1$ structure. IV-simulation for a $2\times2$ reconstruction was unable to arrive at $R_{p}<$ 0.5 for any 2x2 structures we adopted. We also tried As terminated surfaces, also relaxing the atomic positions in the first three layers. This gives much higher Pendry R-factors, R$_{P}\approx$ 0.48 for the $2\times1$ structure and R$_{P}\approx$ 0.42 for the $\sqrt{2}\times\sqrt{2}$ structure. Therefore, based both on robust electrostatic considerations and our quantitative IV LEED study we can conclusively state that the termination surface of the Co-doped Ba122 superconductors cleaved at cryogenic temperatures consists of a half Ba layer coupled to relaxed atomic positions in the near-surface layers.  

The atomic positions from IV-LEED are given in tables in the supplementary information, and clearly indicate significant deviations from the bulk structure. Apart from a small reduction in the inter-planar distances, we also observe buckling of the outermost As-Fe$_{2}$-As block and lateral displacements of the atoms in the first four layers. The Debye temperature of the surface Ba atoms is found to be quite low (60 K), making these Ba atoms relatively mobile. This matches the observation that the fractional spots irreversibly fade out when the sample is warmed to approximately 150 K \cite{massee-PRB-2009}. STM experiments also report destruction of the long range order seen in the topographs by thermal cycling to high temperature \cite{hsieh-arxiv-2008}. Incidentally, this also explains why no clear fractional spots where observed in Ref. \cite{nascimento-PRL-2009}, as those samples were cleaved at room temperature. Seeing as almost all ARPES studies are conducted on cryogenically-cleaved samples, the low T structure reported here is of direct relevance.
\begin{figure}[tbh]

\includegraphics[ height = 8.6 cm]{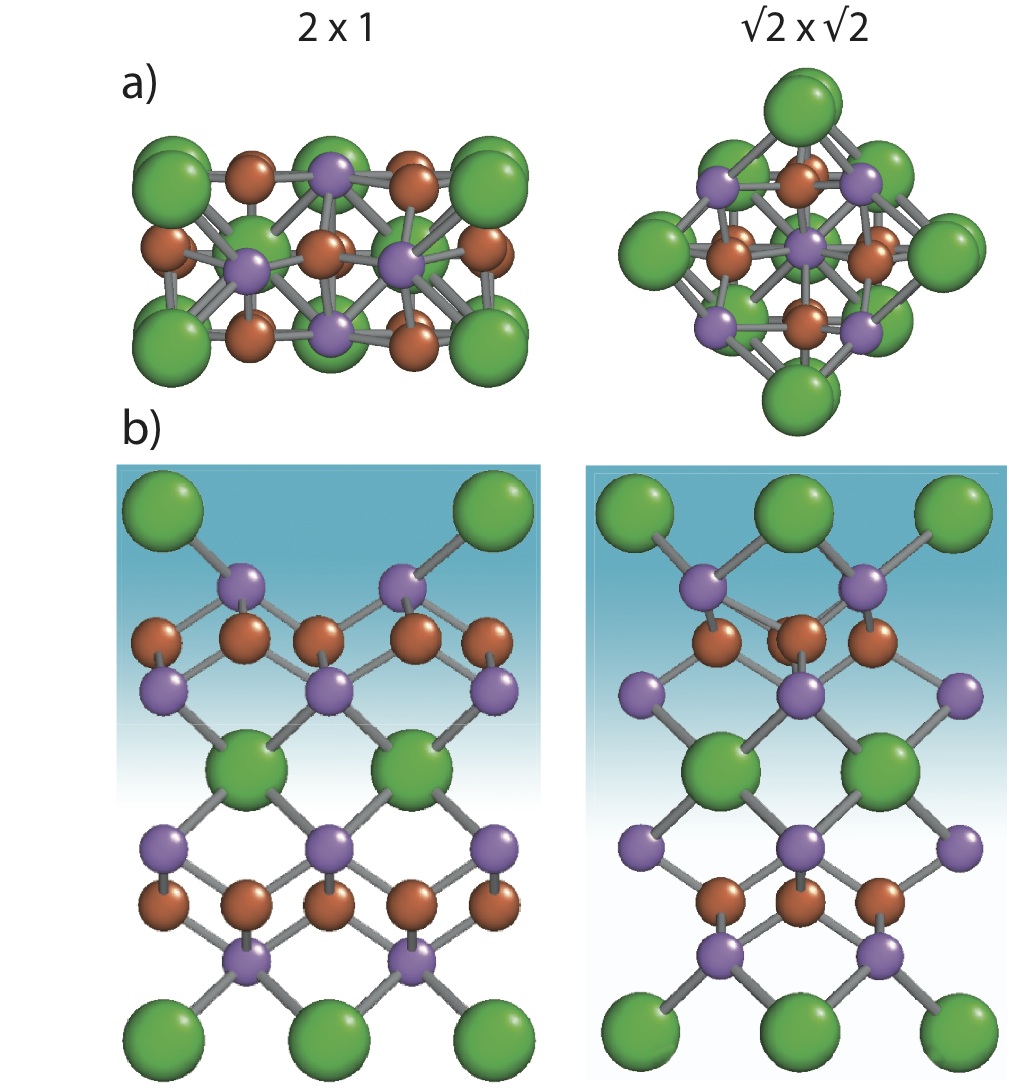}
\caption{(Color online) Top view (a) and side view of half a slab (b) used in the DFT calculations (green: Ba; purple: As; orange: Fe). The blue shaded area in (b) indicates distorted atomic positions, as obtained from the LEED experiments.}
\label{dft_results}
\end{figure}

Now that we have proven the existence of significant (sub)surface reconstructions, the next step is to quantify their effects on the electronic structure. This we do via density functional theory (DFT) calculations for the different surface geometries using the full potential local orbital code (FPLO) \cite{koepernik-PRB-1999} (for details see supplementary information). We simulated the cleavage surface by using $z$-periodic replicas of a finite slab, separated by vacuum (i.e. repeated slab calculations). The structure of the slabs used in the calculations is shown in Fig. \ref{dft_results}(a). In Fig. \ref{dft_results}(a) and (b) we only show the top half of the slab: the other half is identical, but mirrored in the lowest Ba layer shown. To distinguish the effect of the Ba superstructure from the distortions, we repeated the calculations with similar slabs that have all atoms at the bulk positions, but with a $2\times1$ or $\sqrt{2}\times\sqrt{2}$ ordered half-layer of Ba at the surface. To address doping, the virtual crystal approximation (VCA) is employed. 
\begin{figure}[t]
\centering
\includegraphics[width=8.6cm]{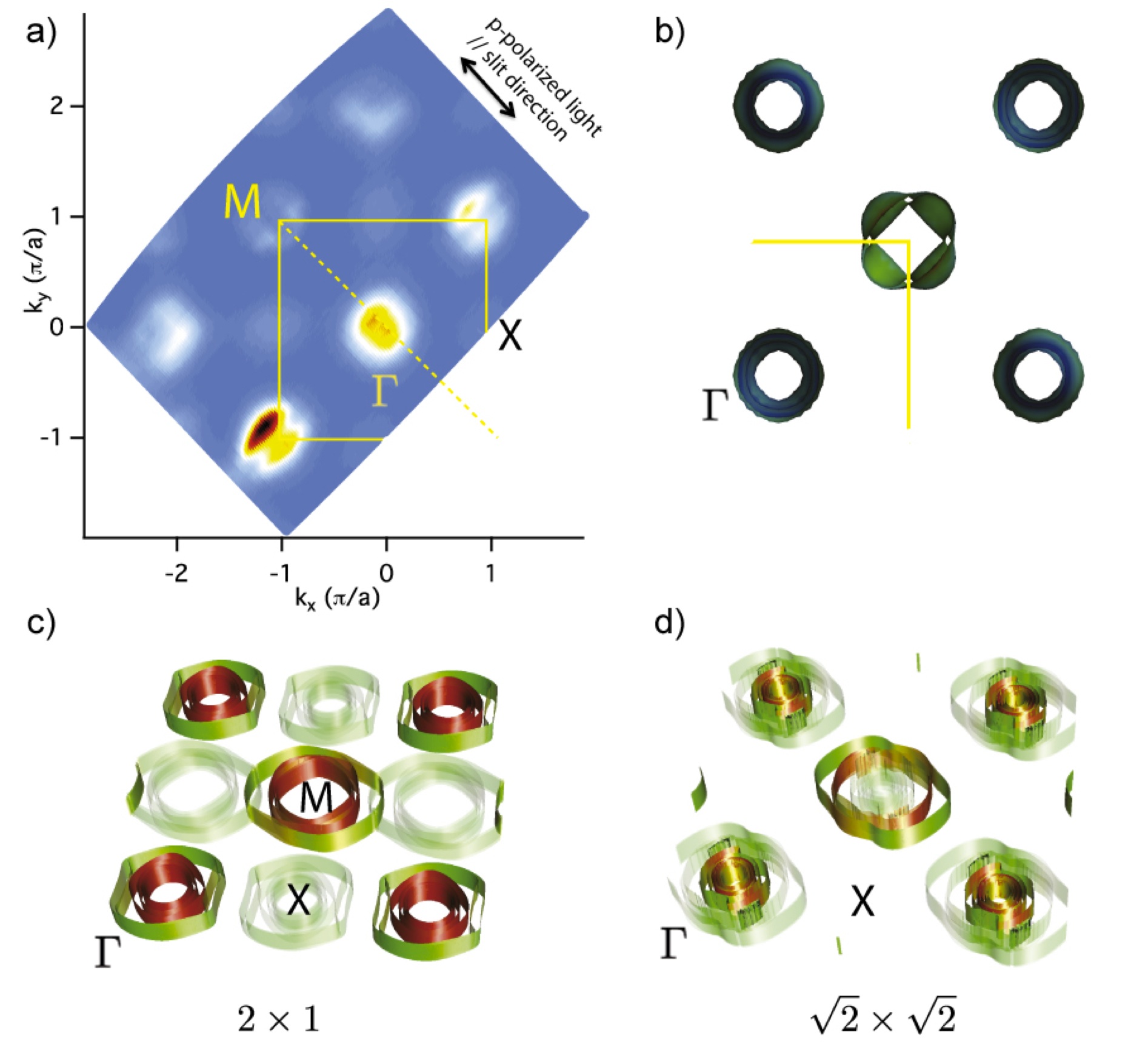}
\caption{(Color online) (a): Fermi surface of BaFe$_{1.83}$Co$_{0.17}$As$_{2}$ measured with 74 eV, p-polarized photons. Indicated in yellow is the 1st Brillouin zone and the $\Gamma$ and $M$ points. The dashed yellow line indicates the cuts shown in Fig. \ref{EDM_comp}. (b): undoped bulk FS from DFT. (c): Unfolded FS for the $2\times1$ termination, from DFT using the relaxed atomic positions and VCA doping of 0.17. Red: bulk; green: surface; brown: mixture of bulk and surface states. (d): same as (c), but for the $\sqrt{2}\times\sqrt{2}$ structure.}
\label{FS_comp}
\end{figure}
The Ba superstructures impose a planar super cell on the whole slab, which has the consequence that all bands, including the bulk bands, are folded back into the smaller Brillouin zone of the super cell. Obviously, this back folding becomes less and less meaningful as the layers become increasingly bulk-like. In the following, we unfold the band structures according to the translational symmetry of the original planar unit cell using the approach of Ref. \cite{ku-PRL-2010} (see supplementary information for details). 
Figure \ref{FS_comp}(a) shows an experimental FS for the x = 0.17 sample measured at 74 eV photon energy. The DFT bulk FS is shown in Fig. \ref{FS_comp}(b). Comparing the two, we note two distinguishing features: (i) a weak FS sheet at the X point in the experimental data and (ii) the fact that the electron FS around M breaks the fourfold symmetry of the bulk crystal. The first feature has also been observed in SrFe$_{2}$As$_{2}$ \cite{hsieh-arxiv-2008}, and disappeared when the sample was heated to 200 K. The same feature is also in the Ca122 data of Ref. \cite{liu-PRL-2009}, making it a ubiquitous feature of the M122 iron pnictide family. Comparing with Fig. \ref{FS_comp}(c), we see that this feature arises from the $2\times1$ and $1\times2$ superstructures. At this stage, we cannot decide which out of the $2\times1$ or $\sqrt{2}\times\sqrt{2}$ structures is responsible for (ii), as both reconstructions lead to significant deviations from four-fold symmetry. We also note that we do not observe the large distortion-induced FS pocket encircling the $\Gamma$ point predicted by the calculations. More extensive variation of the experimental ARPES parameters (photon energy, polarization, geometry) will be necessary to judge to what extent this is a matrix element effect.         
\begin{figure}[t]
\centering
\includegraphics[width=8.6cm]{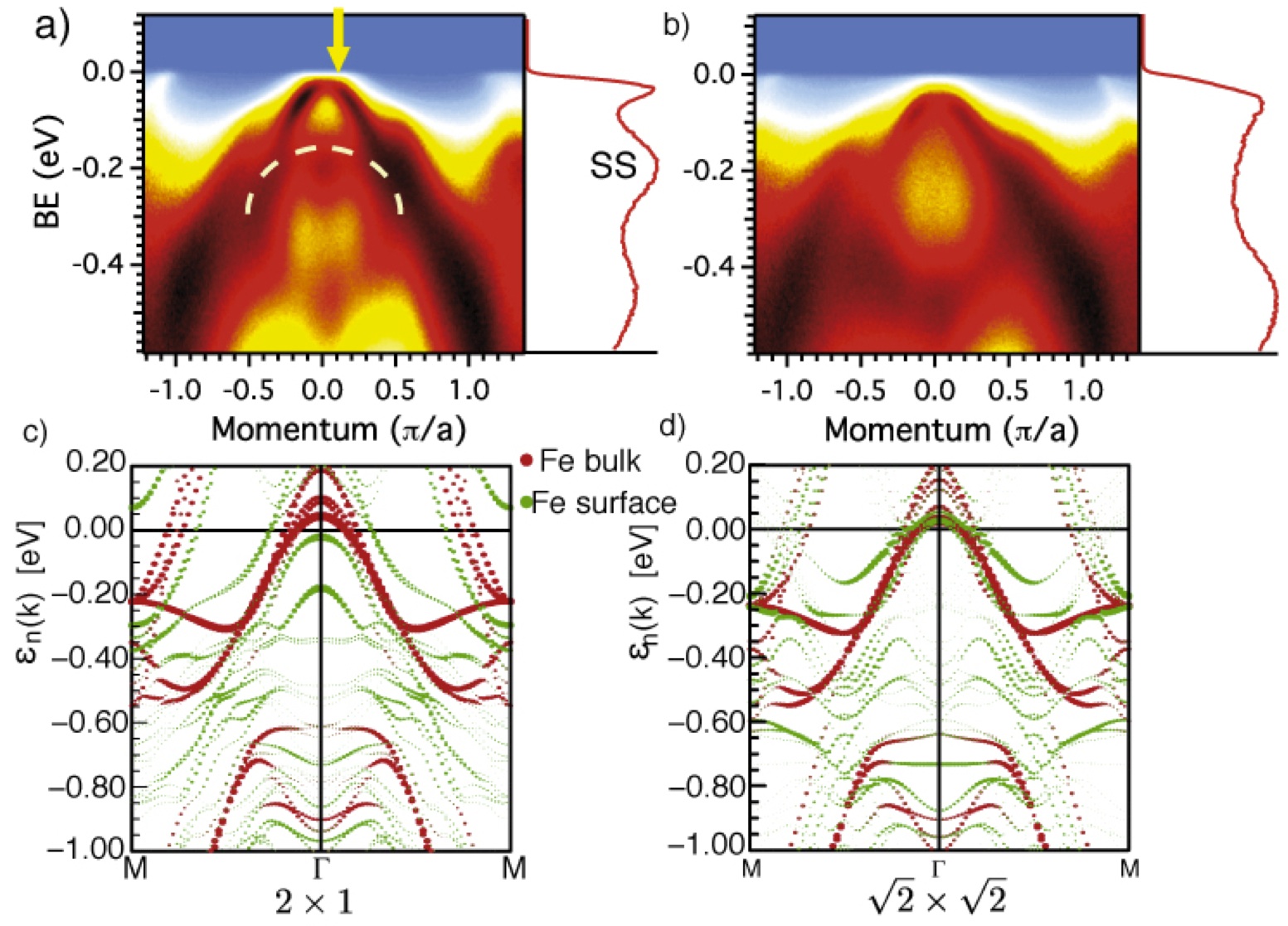}
\caption{(Color online) (a): $I(k,E)$ image taken along the cut indicated in Fig. \ref{FS_comp}(a). On the side we show an EDC taken along the line indicated by the arrow. The dashed curve indicates the surface state (b): same as (a) but taken after cycling in temperature. (c) and (d): calculated band structure for the same momentum window. Note that the energy window is twice as large for the DFT data.}
\label{EDM_comp}
\end{figure}

In Fig. \ref{EDM_comp}, we compare an $I(k,E)$ image taken along the $\Gamma-M$ direction in panel (a) with the bands calculated for the (c) $2\times1$ and (d) $\sqrt{2}\times\sqrt{2}$ distorted structures. As has been observed in all comparisons between ARPES and DFT calculations in the iron pnictides, one can identify most features if one includes a scaling factor renormalizing the energy scale of the DFT data by a factor of 2-3. Apart from this re-scaling, there is good agreement between the calculated and measured spectra. In particular we observe a surface state, indicated in Fig. \ref{EDM_comp}(a) by the dashed yellow curve that gives rise to the central peak in the displayed EDC. Further proof for the identification of this band with surface-related states comes from Fig. \ref{EDM_comp}(b), which is the same cut also measured at 10 K, but taken after cycling the sample temperature to 150 K. Like the fractional spots in the LEED experiments, the surface-related features fade away on the irreversible melting of the ordered Ba superstructure. Comparing the complexity of surface and bulk bands in panels \ref{EDM_comp}(c) and \ref{EDM_comp}(d) with the experimental data, several surface states probably contribute to the experimentally observed surface state. The closeness of the surface and bulk states will inevitably lead to a broadening of the ARPES spectra. It is likely that most of our conclusions carry over to other members of the 122 family, as both termination surfaces have been observed with STM. Ca122 appears to be somewhat special due to the fact that there is a larger energy separation between the $2\times1$ and $\sqrt{2}\times\sqrt{2}$ termination states \cite{gao-PRB-2010}. This is also evident from STM experiments where only $2\times1$ termination is observed \cite{chuang-science-2010}. 

In conclusion, our combined IV-LEED (experiment and theory) and band mapping (experiment and theory) have clearly shown the surface of the Ba122-based high temperature pnictide superconductors to be not only comprised of a 1/2-Ba layer, but also to be significantly distorted with respect to the bulk structure. This gives rise to surface-related bands, which can be clearly identified in ARPES data. These conclusions go a long way to resolving the riddle as to why so much of the published ARPES data on the Ba122 family is so broad: this is due to closely overlapping surface and bulk-derived states. This essentially excludes the use of the Ba122 system for accurate determination of in particular the imaginary part of the self-energies in these superconductors via ARPES. 

We are grateful to H. Luigjes for expert technical assistance. This work is part of the research program of the Foundation for Fundamental Research on Matter (FOM) and VENI program, which are part of the Netherlands Organisation for Scientific Research (NWO). The work in Dresden is carried out under the DFG priority program SPP1458.

\bibliography{paper_citations}

\begin{thebibliography}{27}
\bibitem[{ftn()}]{ftnote}
\bibinfo{howpublished}{Cleavage in the Fe plane is highly unlikely due to the
  much stronger bonding in the Fe-As block}.
\end{thebibliography}

\section{Supplementary material}
\section{Experiments}
Crystals used in the experiments where grown at the University of Amsterdam from self flux. All samples are characterized by resistivity measurements. The critical temperatures are T$_{c}$ = 14 K and 24 K for x=0.10 and 0.17 samples respectively. The x=0.10 sample has a structural/magnetic transition around T$_{st}$ = 43 K and T$_{mag}$ = 35 K.   

The LEED experiments where performed using a modified ErLEED1000 setup from Specs that allows for simultaneous beam-current readout. The spot size for this setup is $<$ 1 mm. The cleaved crystals were aligned with the surface normal axial to the LEED set-up to within 0.2 degrees. The images where captured using a standard CCD camera at 2 eV intervals in primary beam energies. For each set of experiments a dark field image was subtracted from the data. To account for changes in background intensities, we subtracted the background intensity averaged over a square box around the main spot as detailed in \cite{roucka-vac-2002}. 

The ARPES experiments where performed at the SIS-HRES beamline of the Swiss Light Source, using a Scienta R4000 analyzer with a total energy resolution (photon + electron) of 16 meV (at T = 10 K, h$\nu$ = 74 eV) and a momentum resolution better than 0.05 $\pi$/a along the slit.

\subsection{LEED calculations}
We have performed two distinct LEED calculations, as the two coexisting structures ($2\times 1$ and $\sqrt{2}\times\sqrt{2}$) generate different beam sets. Only half-integer beams have been used since the integer beams are created from the intensities of several structures at once. The fractional order beams belong to one of the calculated structures provided that there are no ordered structures with cell size greater than 2 times the bulk unit cell. We have assumed that there are domains in every possible orientation, i.e. four different orientations for both surface structures, and also that there are equal quantities of $2\times1$ and $1\times2$ domains. 
\begin{table}
\caption{The exact coordinates in the $\sqrt{2}\times\sqrt{2}$ termination. The bulk values (z,x,y) are quoted in brackets.}
\begin{tabular}{@{\extracolsep{\fill}}lr@{.}lr@{.}lr@{.}l}
\hline\hline
\multirow{2}{2cm}{Ba}
&$0$&$1717$&$-0$&$1834$&$-0$&$2012$\\
&$0$&$0000$&$0$&$0000$&$0$&$0000$\\
\hline
\multirow{4}{2cm}{As}
&$2$&$0011$&$1$&$5394$&$1$&$6523$\\
&($1$&$8926$&$1$&$9780$&$1$&$9780$)\\
&$2$&$0261$&$1$&$5059$&$-2$&$1909$\\
&($1$&$8926$&$1$&$9780$&$-1$&$9780$)\\
\hline
\multirow{8}{2cm}{Fe}
&$3$&$2869$&$3$&$9686$&$-2$&$1561$\\
&($3$&$2363$&$3$&$9559$&$-1$&$9780$)\\
&$3$&$3294$&$3$&$7308$&$1$&$7104$\\
&($3$&$2363$&$3$&$9559$&$1$&$9780$)\\
&$3$&$3778$&$5$&$7287$&$-0$&$3662$\\
&($3$&$2363$&$5$&$9339$&$0$&$0000$)\\
&$3$&$3832$&$1$&$8588$&$-0$&$3138$\\
&($3$&$2363$&$1$&$9780$&$0$&$0000$)\\
\hline\hline
\label{sqrt2x2}
\end{tabular}
\end{table}
Since all orientations are observed with roughly equal frequency in STM experiments on a single cleave, we average the experimental IV curves of symmetry equivalent spots (eg. (1,1),(1,-1),(-1,1),(-1,-1) are averaged). In the calculations no symmetry is enforced and equivalent spots have been averaged. As a result the calculated LEED patterns have four fold symmetry even if the surface structures do not have this symmetry. The bulk unit cell of this sample is tetragonal so that the surface unit cells are exactly $2\times 1$ and $\sqrt{2}\times\sqrt{2}$. Resistivity measurements on the crystals used show no sign of the tetragonal to orthohombic transition for the x = 0.17 sample for which the calculations where performed. The tetragonal two dimensional unit cell has been assumed to be of size 3.9624 $\text{\AA}$ $\times$ 3.9624 $\text{\AA}$. This is an average of quoted values \cite{rotter-PRB-2008, sefat-PRL-2008}. The LEED calculations were performed using the Barbieri/Van Hove SATLEED package \cite{satleed}. The 12 phase shifts needed in the LEED calculation were calculated with the Barbieri/Van Hove phase shift package \cite{Barbieri}. Temperature effects were calculated within the SATLEED code by multiplying each atom's scattering amplitude by a Debye-Waller factor. Pendry R-factors \cite{pendry-JPC-1980} were used to measure the level of agreement between measured and calculated IV spectra. Statistical errors in analysis were estimated with Pendry RR-factors \cite{pendry-JPC-1980}. We have used the averaged T-matrix approximation (ATA) \cite{Beeby-PRB-1964} to take into account the Co-doping. In this approximation the Fe/Co -atoms are replaced by an averaged atom, with a scattering amplitude which is a weighted average of the scattering amplitudes of Fe and Co. The results would not differ much if only Fe would be used in this place as the Fe and Co are very similar scatterers and the amount of Co is not too big. During the calculations the surface geometry and the complex inner potential were optimized. The Debye temperatures have been optimized for each element. The Ba surface-atoms have a different Debye temperature from the ones deeper in the bulk. The Debye temperature for Fe is optimized for Fe solely. Because the $T_D$ of Co and Fe are close to each other and the amount of Co is small, the Co doping should not affect the optimized $T_D$ for Fe.

The optimized atomic positions obtained for the two different reconstructions are collected in table \ref{sqrt2x2} for the $\sqrt{2}\times\sqrt{2}$ terminated surface and in table \ref{1x2} for the $2\times1$ termination.
\begin{table}
\caption{The exact coordinates in the $2\times1$ termination. The bulk values (z,x,y) are quoted in brackets.}
\begin{tabular}{@{\extracolsep{\fill}}lr@{.}lr@{.}lr@{.}l}
\hline\hline
\multirow{2}{2cm}{Ba}
&$0$&$2730$&$-0$&$0243$&$-0$&$3503$\\
&$0$&$0000$&$0$&$0000$&$0$&$0000$\\
\hline
\multirow{4}{2cm}{As}
&$2$&$0895$&$1$&$8537$&$1$&$5570$\\
&($1$&$8926$&$1$&$9780$&$1$&$9780$)\\
&$2$&$1125$&$5$&$7112$&$1$&$7153$\\
&($1$&$8926$&$5$&$9339$&$1$&$9780$)\\
\hline
\multirow{8}{2cm}{Fe}
&$3$&$3173$&$1$&$9307$&$-0$&$1066$\\
&($3$&$2363$&$1$&$9780$&$0$&$0000$)\\
&$3$&$3266$&$6$&$0300$&$-0$&$3240$\\
&($3$&$2363$&$5$&$9339$&$0$&$0000$)\\
&$3$&$3885$&$3$&$7040$&$1$&$9820$\\
&($3$&$2363$&$3$&$9559$&$1$&$9780$)\\
&$3$&$4274$&$-0$&$1746$&$2$&$1709$\\
&($3$&$2363$&$0$&$0000$&$1$&$9780$)\\
\hline\hline
\label{1x2}
\end{tabular}
\end{table}
Compared to the bulk structure \cite{rotter-PRB-2008}, substantial lateral changes occur in the three topmost layers [table \ref{geom}]. The Fe-layer approximately 3 $\text{\AA}$ below the surface Ba-atoms still has lateral changes of the order 0.4 $\text{\AA}$ in both terminations. However, since LEED is more sensitive to geometric parameters normal to the surface than lateral parameters, the error margins on the quoted lateral movements are quite big. The estimated errors in lateral geometric results are given for the direction joining the optimized atomic position and the bulk position. This means that the Pendry RR-factors are calculated along the line from bulk position to the best fit position. The errors are calculated one at a time, while the other atoms reside on the best fit positions. On doing so, all lateral coordinates have error margins of the order $\pm$ 0.2 $\text{\AA}$ and all coordinates normal to the surface have error margins of the order $\pm$ 0.04 $\text{\AA}$. Apart from the lateral movements the first As-Fe$_{2}$-As block below the terminating Ba layer becomes buckled and the interlayer distances dimisnish.  
\begin{table}
\caption{The geometric structure of the BaFe$_{2-x}$Co$_{x}$As$_{2}$ surface for the two different surface structures. The distances dz$_{Ba-As}$, dz$_{As-Fe}$ and dz$_{Fe-As}$ are measured between the highest and lowest atom in neighboring sheets. Average error bars on the inter-planar distances are $\pm$ 0.04, while those on the lateral movements are $\pm$ 0.18. All values are in $\text{\AA}$. }
\begin{tabular}{lccc}
\hline\hline
&\multicolumn{1}{c}{bulk}&\multicolumn{1}{c}{$\sqrt{2}\times\sqrt{2}$}&\multicolumn{1}{c}{$2\times 1$}\\
$dz_{Ba-As}$&\multicolumn{1}{c}{1.89}&\multicolumn{1}{c}{1.83}&\multicolumn{1}{c}{1.82}\\
$dz_{As-Fe}$&\multicolumn{1}{c}{1.34}&\multicolumn{1}{c}{1.26}&\multicolumn{1}{c}{1.20}\\
$dz_{Fe-As}$&\multicolumn{1}{c}{1.34}&\multicolumn{1}{c}{1.22}&\multicolumn{1}{c}{1.20}\\
\\
\textit{buckling}&&&\\
As-layer&&\multicolumn{1}{c}{$0.03$}&\multicolumn{1}{c}{$0.02$}\\
Fe-layer&&\multicolumn{1}{c}{$0.10$}&\multicolumn{1}{c}{$0.11$}\\
\\
\textit{lateral shifts}&&&\\
Ba-layer&&\multicolumn{1}{c}{$0.27$}&\multicolumn{1}{c}{$0.35$}\\
As-layer&&\multicolumn{1}{c}{$0.55,0.52$}&\multicolumn{1}{c}{$0.44,0.34$}\\
Fe-layer&&\multicolumn{1}{c}{$0.18,0.35$}&\multicolumn{1}{c}{$0.12,0.34$}\\
&&\multicolumn{1}{c}{$0.42,0.34$}&\multicolumn{1}{c}{$0.26,0.26$}\\
\hline\hline
\label{geom}
\end{tabular}
\end{table}

When integer beams and half integer beams belonging to the $\sqrt{2}\times\sqrt{2}$ termination are taken simultaneously into the calculation, one gets a structure that is nearer to the bulk structure but the total Pendry R-factor increases to 0.32 from the 'only half integer beams' -value 0.19. The dz-values grow towards the bulk values (1.84 $\text{\AA}$,1.32 $\text{\AA}$,1.27 $\text{\AA}$) and buckling diminishes (0.01$\text{\AA}$ for the As layer, 0.07$\text{\AA}$ for the Fe layer). 

The thermal movement of the atoms is taken into calculation by fitting Debye temperatures. For $2\times 1$ termination, the best fit value for Ba is $60 K$ which is quite low compared to the tabulated bulk value of $110 K$ \cite{kittel}. For $\sqrt{2}\times\sqrt{2}$ termination the best fit value is $70 K$. These values are lower than the usual bulk value divided by $\sqrt{2}$ for surface atoms which would give $78 K$. It means that the surface Ba is very mobile which is to be expected since it can form different terminations at low temperatures. For As the tabulated value for the Debye temperature is $282 K$. We find that the Debye temperature is much higher than that 
but the sensitivity in the simulations to the $T_D(As)$ is so low that no precise value can be given. The $T_D$ for Fe is $675 K$ for $\sqrt{2}\times\sqrt{2}$ termination and $425 K$ for $2\times 1$ termination. These values are of the same order of magnitude as the tabulated value for Fe ($470 K$).

\subsection{LDA calculations}
\begin{figure}[t]
\begin{centering}
\includegraphics[width=8.6cm]{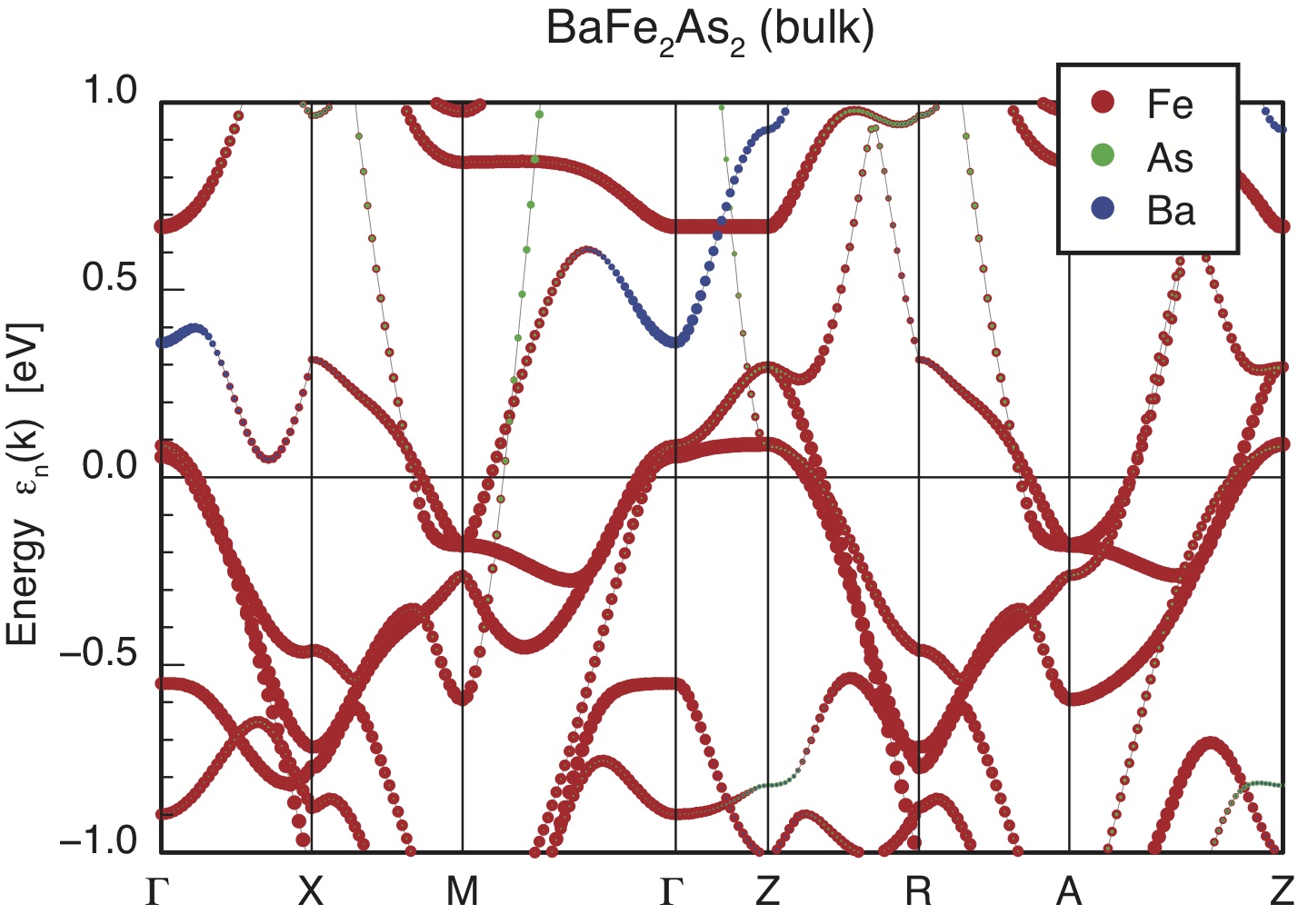}
\end{centering}
\caption{\label{fig:bct-bulk-BS}The bulk bandstructure of
undoped BaFe$_{2}$As$_{2}$.  The band structure is colored according to
orbital weights.}
\end{figure}

We use the experimental lattice parameters and Wyckoff positions and focus on temperatures above the spin density wave transition, which allows for non-spin polarized calculations. Given, that we do not relax the lattice we also do not have to worry too much about the choice of the exchange and correlation functional used in the calculations. Experience shows, that although lattice degrees of freedom are notably influenced by the choice of the exchange and correlation functional, band structures are quite insensitive to the choice (given the same lattice structure).

In order to correctly treat the boundary condition of the Poisson equation we used a symmetric slab, with a $\sigma_{z}$-mirror plane in the center of the slab. In that way we prohibit the building up of artificial surface dipoles. The self consistent calculations used a $6\times6\times1$ $k$-mesh and the tetrahedron method for the Brillouin zone integration. Note, that the slabs have larger in-plane unit cells than the bulk setup, due to the surface Ba superstructures. That makes the number of $k$-points taken sufficient. Relativistic effects were treated on a scalar relativistic level. We use repeated slab calculations in which the slabs are separated by vacuum. We made sure that the vacuum between the slab is wide enough to decouple the replicas. In a slab calculation the $k_{z}$-dispersion is basically replaced by a projection of the band structure onto the $k_{x},k_{y}$-plane. Due to the restriction of the calculations to a rather small finite slab we will only get a limited number of Fermi surfaces in the region of the projected $k_{z}$-dispersion. In our case we have two bulk and two surface FeAs triple layers, which is of course not enough to show all of the $k_{z}$-projected bands. However, we will get representatives of the bulk bands in the correct regions in $k$-space, which can be compared to pure surface bands.
\begin{figure}[t]
\noindent \begin{centering}
\includegraphics[width=8.6cm]{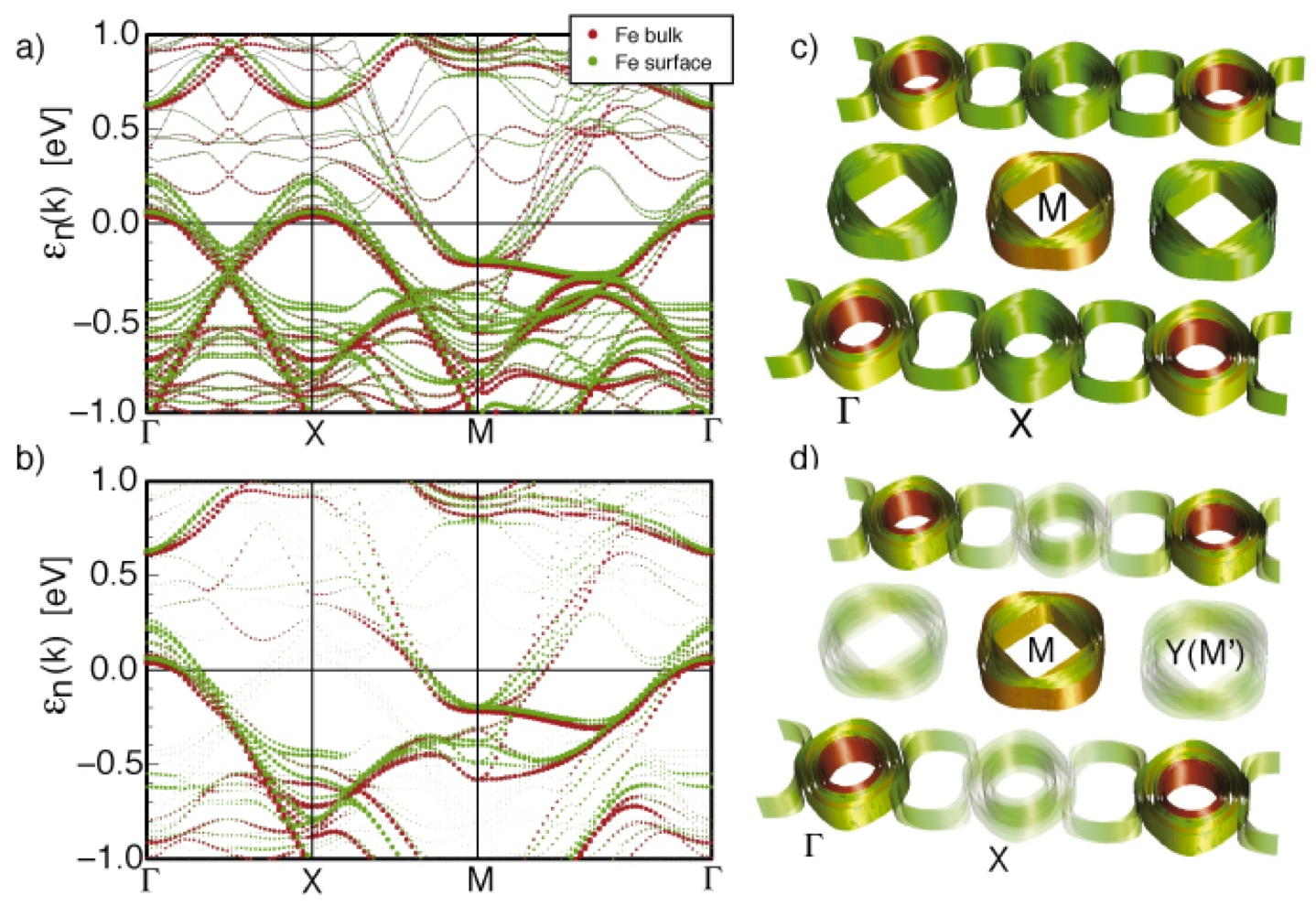}
\par\end{centering}

\caption{\label{fig:unfolding}Demonstration of the band unfolding technique
for the doped ordered $2\times1$ slab. a) Band structure without
unfolding. b) unfolded band structure. c,d) Fermi surface without
and with unfolding. }

\end{figure}

There are several indicators for convergence of a finite slab to a finite slab with bulk like electronic structure far away from the surface. The first is, that the inner most layers of the finite slab
have a population analysis (and hence band structure/density) unchanged from the bulk calculation. The slabs we used have four Fe$_{2}$As$_{2}$ triple layers and it turns out that the two inner triple layers agree in their population numbers with the bulk results within a few hundredth of electrons. The second issue is that a charged surface will locally change the chemical potential, which manifests itself in band shifts of the surface bands compared to the bulk bands. This difference of the Fermi level between the surface and the bulk will gradually vanish going from the surface into the bulk. Hence, another parameter to be converged is the Fermi level of the bands belonging to the inner
most layers. In our case the surface charging is less pronounced than in the 1111 case \cite{eschrig-PRB-2010} since the formation of Ba superstructures restores the local stoichiometric chemical composition, which in turn makes our four triple layer slab super cells large enough. 

\begin{figure*}[t]
\noindent \begin{centering}
\includegraphics[width=17cm]{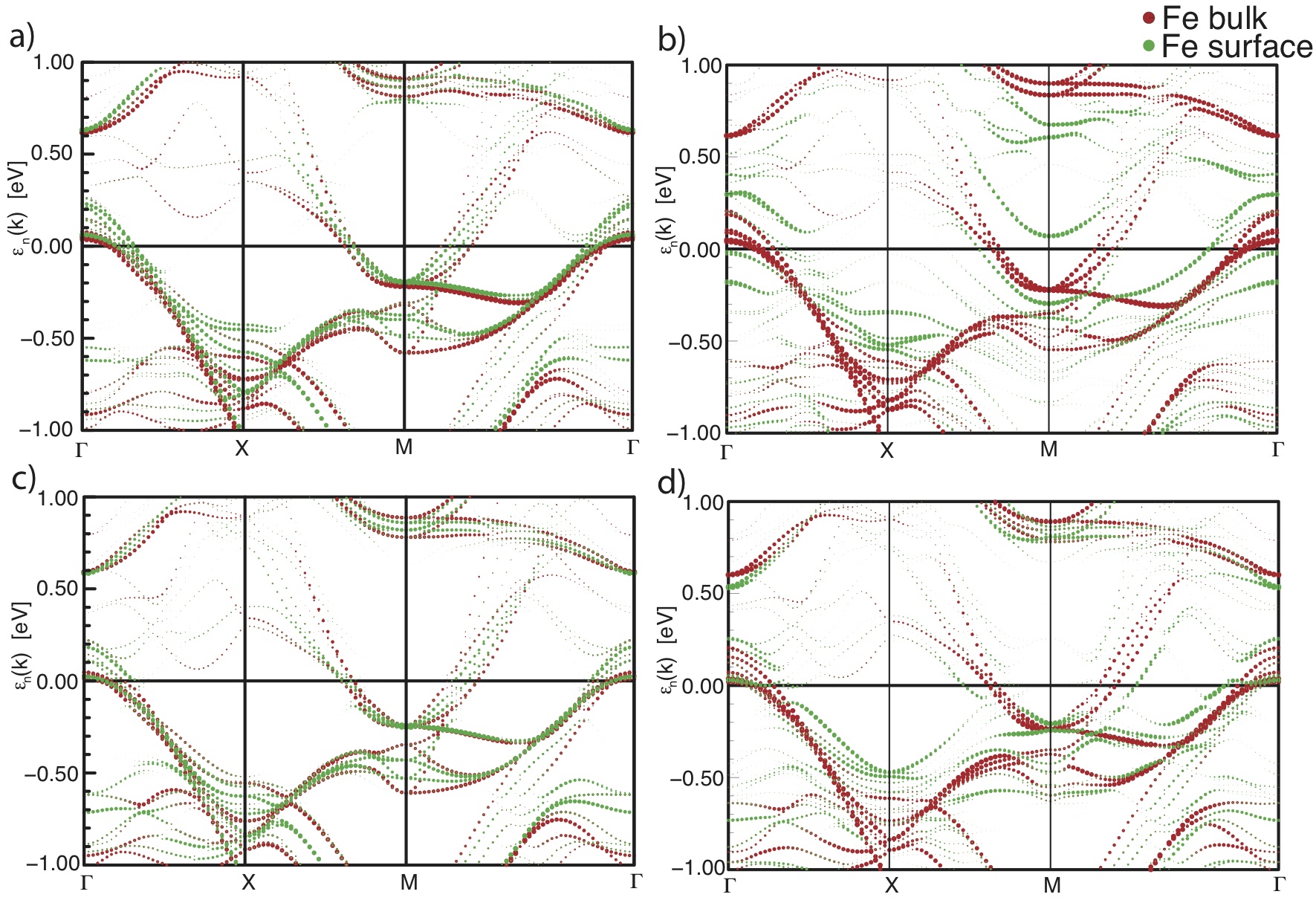}
\par\end{centering}

\caption{\label{fig:dft_results}(Color Online) Comparison of slabs with superstructure only ((a),(c)) and with distortions in the Ba layer and As-Fe$_{2}$-As block ((b),(d)). The top (bottom) panels are for the $2\times1$ ($\sqrt{2}\times\sqrt{2}$) superstructure. }

\end{figure*}
To illustrate the change in occupations, we focus on the Mullikan type population analysis obtained by projection onto the local orbitals. Due to hybridization and overlap effects the occupation numbers do not reflect one to one the chemical valency, however trends can be obtained quite easily. In the bulk the Ba misses one electron, while the two As per formula unit each have $0.55$ excess electrons (while the difference to neutrality sits on the two Fe atoms). In the slab calculations the single surface Ba misses only $0.49$ electrons, while the top most (sub surface) As layer now has $0.33$ excess electrons on average. In the surface superstructures considered here every second surface Ba atom is missing while the FeAs triple layers have two Fe$_{2}$As$_{2}$ formula units. Counting Ba atoms correctly reveals that each four-Fe triple layer must be combined with one Ba from above the layer and from below the layer in order to obtain the stoichiometric BaFe$_{2}$As$_{2}$ formula unit. So the two formula units per unit cell of the surface layers have $1.5$ missing electrons, which when compensated by the top most As layer leads to an As occupation $\frac{3}{8}$, which is quite close to what we obtained. The situation is more or less the same for all slabs considered, although there are some variations in how much the charge compensation sits only on the As layer under the surface compared to also sitting to a lesser degree on the As layer below the topmost Fe layer.

In a recent publication \cite{ku-PRL-2010} a technique for band un-folding was described. The original idea is based on Wannier functions, however as is pointed by the authors a local orbital basis is a suitable starting point as well. In fact it turns out that the technique is even simpler in a local basis code. The usual {}``fat-band'' plots are obtained by projecting the Bloch states onto local orbitals of a certain type and symmetry, for instance Fe $3d$ orbitals. One can spin the symmetry selective properties of such a projection a bit further by including translational symmetry, which is achieved by using Bloch sums of these local orbitals as projectors. If the phase factors of the Bloch sums are chosen such that they become representations of a certain translational symmetry, the resulting band weights will show destructive interference for states, which are replicas of the bands according to the chosen translational symmetry. Ref \cite{ku-PRL-2010} points out that the Bloch-sum orbital weights as obtained by this approach contain some of the matrix element effects, which enter real life measurements such as ARPES. We implemented this technique into FPLO and will use it in what follows in order to simplify the discussion. The trick allows not only the unfolding of bands but naturally also includes projection on certain layers of the slab. 

Figure \ref{fig:bct-bulk-BS} shows the well known band structure, while the Fermi surface of undoped bulk BaFe$_{2}$As$_{2}$ for T $>$ T$_{st}$ $>$ T$_{mag}$ is shown in Fig. 4(b) of the main paper. The line thickness in Fig. \ref{fig:bct-bulk-BS} represents the orbital character of the corresponding bands. The main contributions to the bands around the Fermi level comes from the Fe 3d orbitals (red), with small admixtures of As 4p (green). There is a special feature of the Ba122 compound between 0 and 400 meV above the Fermi level along the line $\Gamma$X, which is due to a Ba 5d band. Since, the iron character is so dominating we will use the Fe 3d orbitals in the following, whenever projections onto surface and bulk layers are needed. 

We have performed calculations for four different slab geometries, i.e. the two distorted slabs as shown in Fig. 3 of the main paper  and the corresponding undistorted (ordered) slabs, which have the $\sqrt{2}\times\sqrt{2}$ and $2\times1$ planar Ba surface superstructures but bulk positions and distances otherwise. 

The consequences of the unfolding technique are illustrated in Fig. \ref{fig:unfolding}, which shows the band structure (BS) and Fermi surfaces (FS) for the undistorted $2\times1$ slab along the symmetry lines of the simple planar BZ of the Fe$_{2}$As$_{2}$ unit cell. In the upper panel the standard bandstructure calculation result without unfolding is shown. The backfolding due to the $2\times1$ super cell is clearly seen along the line $\Gamma$X. This is also illustrated by the FS in Fig. \ref{fig:unfolding}(c), where replicas of the FS sheets around $\Gamma$ are found around the X-point. The green sheets are predominantly of Fe surface layer character (the iron layer closest to the surface), while the red (brown) sheets are predominantly of bulk (mixed) character. It is easy to understand that the backfolded bands around X have nearly exclusively surface character, since the effect of the Ba surface superstructure on the electronic structure decays with increasing layer depth. A similar discussion holds for the replica of the bands around the M-point, which appear at Y(M$^{\prime}$). In Fig. \ref{fig:unfolding}(b) and \ref{fig:unfolding}(d) the results after unfolding are shown. We chose the same color code. Most notably, the back folded bands around the X-point fade away due to the destructive interference resulting from the unfolding projection. In Fig. \ref{fig:unfolding}(d) we enhanced the contrast of the these replicas for better visibility. 

We now discuss the main differences arising from the distortions in the surface and first few subsurface layers. The bandstructures corresponding to the undistorted and distorted $2\times1$ and $\sqrt{2}\times\sqrt{2}$ superstructures are shown in Fig. \ref{fig:dft_results}. A striking difference between the two sets of calculations is that the distortions lifts the degeneracy between bulk and surface bands.
\begin{figure}[t]
\noindent \begin{centering}
\includegraphics[width=1\columnwidth]{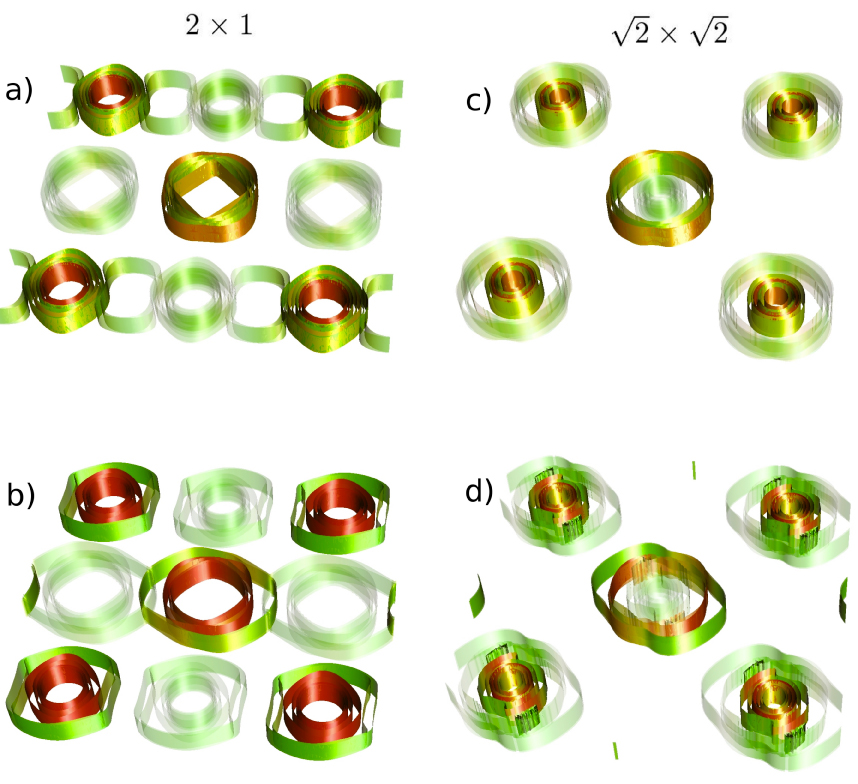}
\par\end{centering}

\caption{\label{fig:fscomparison}Unfolded Fermi surfaces of the $2\times1$
(left) and $\sqrt{2}\times\sqrt{2}$ (right) Ba surface layer super
structures. Upper (lower) panel: with undistorted (distorted) surface and sub-surface layers.
The unfolding technique was used. Green: bands from the Fe-surface
layer. Red (brown): bulk (mixed) states.}

\end{figure}
In Fig. \ref{fig:fscomparison} we show the summary of the Fermi surfaces for various Co-doped slabs. The upper panel shows the result for the undistorted slabs (with $2\times1$ and $\sqrt{2}\times\sqrt{2}$ Ba termination layers) and in the lower panel including the distortions of the Ba, and underlying As-Fe$_2$-As block. The left panel belongs to the $2\times1$ and the right panel to the $\sqrt{2}\times\sqrt{2}$ slabs. A general result of our calculations is that the virtual crystal approximation at the Fe-site leads to a rather rigid band shift in the energy window of $\pm$1 eV around the Fermi level. This is in contrast to the strong non-rigid band behavior upon VCA doping at the Ba-site \cite{singh-PRB-2008} and can be understood as the effect of the increased nuclear charge at the VCA Fe-sites, which changes the effective electrostatic potential for the iron layers. The main effect of VCA Co-doping hence is an increase of the Fermi radii around the M-point and a decrease of the Fermi radii around $\Gamma$. This trend is seen when comparing the FS of the undoped compound in Fig. 4(b) of the main paper with Fig. \ref{fig:fscomparison}(a) and \ref{fig:fscomparison}(c). The conclusion that can be drawn from Fig. \ref{fig:fscomparison} is that the undistorted slabs (panel (a) and (c)) strongly resemble the $k_{z}$-projected bulk Fermi surfaces, except for the replica FS sheets in panel a, which distinguish between the $\sqrt{2}\times\sqrt{2}$ and the $2\times1$ surface superstructures. Around the M-point in the center of the pictures the two $90^{\circ}$ rotated ellipses, known from the bulk BS, are identifiable. (Note, that not all projected bulk bands are present in these pictures due to the finite slab size.) The comparison of the Fermi surfaces of the undistorted (upper panels a and c) and the distorted (lower panels b and d) slabs shows that there is a strong deviation of (especially) the surface bands from the tetragonal symmetry due to the symmetry breaking caused by the buckling and lateral distortions of the first few surface layers. This is more pronounced for the $\sqrt{2}\times\sqrt{2}$ structure [Fig. \ref{fig:fscomparison}(d)].

\end{document}